\documentclass[10pt]{article}
\usepackage{hyperref,graphicx}   
\usepackage{amsfonts}
\usepackage{amssymb}
\usepackage{amsbsy}
\usepackage{amsmath}
\usepackage{latexsym}
\usepackage{bm}
\usepackage{color}
\usepackage{wasysym}
\usepackage{mathbbol}

\begin{document}
\title{{\bf{\Large  Field theory in Rindler frame and more on the correspondence with thermal field theory formalisms}}}
\author{
 {\bf {\normalsize Dipankar Barman}$
$\thanks{E-mail: dipankar1998@iitg.ac.in}},\, 
 {\bf {\normalsize Bibhas Ranjan Majhi}$
$\thanks{E-mail: bibhas.majhi@iitg.ac.in}}\\
 {\normalsize Department of Physics, Indian Institute of Technology Guwahati,}
\\{\normalsize Guwahati 781039, Assam, India}
\\[0.3cm]
}

\maketitle

\begin{abstract}
Considering two accelerated observers with same acceleration in two timelike wedges of Rindler frame we calculate the Feynman-{\it like} propagators for a real scalar field in a thermal bath with respect to the Minkowski vacuum. Only the same wedge correlators are symmetric under the exchange of the {\it real} thermal bath and Unruh thermal bath, while the cross-wedge ones are not. Interestingly, they contain a cross term which is a collective effects of acceleration and thermal nature of field. Particularly the zero temperature description along with {\it no analytic continuation} between coordinates in right and left Rindler wedges, as expected, corresponds to usual thermofield-double formalism. However, unlike in later formulation, the two fields are now parts of the original system. Moreover it bears the features of a spacial case of closed-time formalism (CTP) where the Keldysh contour is along the increasing Rindler time in the respective Rindler wedges. Interestingly, we observe a new feature that the analytic continuation between the wedges provides the two more spacial cases of CTP. Hence Rindler-frame-field theory seems to be a viable candidate to deal thermal theory of fields and may illuminate the search for a bridge between the usual existing formalisms.  
\end{abstract}

\section{Introduction}
 In zero-temperature limit the conventional quantum field theory (QFT) predictions match exceptionally well with the existing experiments. However, many real-world systems can not be well approximated as zero-temperature ones. The study of hot quark-gluon plasma \cite{cite-key:quark, Laine:2016hma}, thermal phase transitions \cite{Witten:1998zw, Laine:2016hma}, cosmological inflation \cite{PhysRevD.31.1225, RevModPhys.57.1, PhysRevLett.67.793, PhysRevD.54.2519, BERERA2000666, Laine:2016hma}, and the evolution of a neutron star \cite{Laine:2016hma} requires a sophisticated formalism of thermal field theory. There are various approaches to deal the thermal nature of a system. Usually the thermal ensemble average of any operator is done with the thermal density operator $\rho(\beta)=e^{-\beta\mathcal{H}}$ where $\beta$ and $\mathcal{H}$ are the inverse temperature and Hamiltonian of the system, respectively. A significant development in understanding the theory of fields in thermal equilibrium is made by Matsubara \cite{Matsubara:1955ws}, which is known as Matsubara formalism or imaginary time formalism. In this formalism, time takes imaginary values from $0$ to $-i\beta$ and the limit variable $\beta$ plays the role of the system's inverse temperature. The thermal Green function is then evaluated from that of the zero temperature field theory through an infinite series-sum. The thermofield double (TFD) formalism \cite{DasFTFT, Cottrell:2018ash, Azizi:2023smk} is another one in which thermal vacuum state is defined to mimic the ensemble average. In this construction, one needs to add a fictitious (tilde) system along with the original system, which is identical to the target system, except it follows conjugation rules \cite{1987PhR...145..141L, DasFTFT}. The TFD state (expressed as $|0,\beta\rangle$) is a superposition state of the combined eigenstates of two identical systems ($|n\rangle\otimes|\tilde{n}\rangle$) and is connected to the zero-temperature vacuum of the combined system by a Bogoliubov transformation. Such situation then corresponds to four thermal propagators. These are, the Feynman propagator for the original system, the anti-Feynman propagator for the tilde system, and two cross propagators for combination of fields in both original and tilde systems. These propagators form a $2\times2$ matrix comprising zero-temperature and thermal parts of the propagators. 
 This formalism plays a vital role in understanding the time-evolution of entanglement entropy \cite{Hartman:2013qma, Chapman:2018hou, PhysRevLett.127.080602, Dadras:2019tcz}, scrambling and quantum chaos \cite{Shenker:2013pqa, Shenker:2013yza, Roberts:2014isa, delCampo:2017ftn}, firewalls \cite{Almheiri:2013hfa, Maldacena:2013xja, Papadodimas:2012aq}, ER=EPR \cite{Maldacena:2013xja, Susskind:2014yaa}, and emergent spacetime \cite{VanRaamsdonk:2010pw}, etc.

Closed-time path (CTP) formalism, also known as ``in-in'' or real-time formalism (mainly based on Schwinger-Keldysh formulation) is another one to deal thermal theory of fields \cite{doi:10.1063/1.1703727,  Keldysh:1964ud}. One particular choice of the Keldysh contour (see Fig. \ref{fig:contour}) has two real parts ($C_{+}$ and $C_{-}$) and two imaginary parts ($C_3$ and $C_4$). The assigning of two physical fields on two real time axes ($\phi_+$ on $C_+$ and $\phi_-$ on $C_-$) provides four components of propagator and therefore again can be represented by a $2\times 2$ matrix. The branch $C_{3}$ has imaginary values from $0$ to $-i\sigma$ with $0\leq \sigma \leq \beta$. It has been observed that the choice $\sigma=\beta/2$ provides the thermal propagators which are identical to those obtained in TFD with the identification -- original and tilde fields in TFD as $\phi=\phi_+$ and $\tilde{\phi}=\phi_-$, respectively. The time in the $C_{-}$ branch runs in the opposite direction as compared to the $C_{+}$ branch. This is analogous to the conjugation rule for the tilde fields in TFD. In spite of such similarities, these two formalism are not physically equivalent. In TFD the tilde field is fictitious and both the fields commute with each other. Whereas, the fields in CTP have physical existence and their commutation relations are not such trivial one. Moreover in interacting theory these formalisms behave differently. 

Nonetheless, there are certain advantages in the respective formulations. As described, TFD is capable of describing the thermal-vacuum state, while CTP dose not. But the later one deals with two physically existing fields. Then naturally one might be wondering whether there is any other way of describing thermal field theory which can provide a middle root such that those issues can be tackled in a better way. We feel through our present investigation that the field theory in Rindler frame can be one such candidate. Let us now provide the motivation for such possibility.
It is well known that an accelerated observer always gets casually restricted to a quadrant of the Minkowski spacetime. The null paths passing through the origin act as the Rindler horizons for the observer. These paths divide the whole spacetime into four parts: left (L), right (R), past (P), and future (F) Rindler wedges. In this case two timelike regions are possible, namely the right Rindler wedge (RRW) and the left Rindler wedge (LRW). The regions RRW and LRW are causally disconnected, and their coordinate times run in the opposite directions. Also existence of two fields ($\phi^R$ and $\phi^L$) naturally appears in these two accelerating frames. These features are similar to CTP formalism. Moreover, since the respective frame see the Minkowski vacuum as thermal bath (known as Unruh effect \cite{Unruh:1976db,Unruh:1983ms,book:Birrell,Crispino:2007eb}), it is possible that the thermal vacuum state in TFD formalism can be mimicked by the Minkowski vacuum while expressing this as a linear combinations of composite Fock states of the two Rindler fields. Therefore it seems that the field theory in Rindler frames may be a natural candidate to describe thermal theory of fields.

In this paper we consider two observers which are accelerating with same constant acceleration $a$ in RRW and LRW, respectively. For generality, the Rindler fields (which are taken to be real scalar ones) are considered as thermal at inverse temperature $\beta_0$ and the spacetime is taken to be $(1+1)$-dimensional. All the possible combinations for Feynman propagators, namely $G_{RR}$ , $G_{LL}$, $G_{RL}$ and $G_{LR}$, are being calculated through ensemble averaging procedure with respect to the vacuum of the Minkowski observer. Two particular situations are being considered -- without as well as with applying analytic continuation among the coordinates between RRW and LRW. We observe that $G_{RR}$ and $G_{LL}$ individually consists of four parts -- usual non-thermal, thermal due to $\beta_0$, Unruh thermal due to acceleration at inverse temperature $\beta_U = 2\pi/a$ and a cross term due to both $\beta_0$ as well as $\beta_U$. These two correlators are symmetric under the exchange of $\beta_0\leftrightarrow\beta_U$. Whereas other two consist of a pure Unruh thermal and a cross-term. These do not carry the symmetric property. The cross contribution can be  interpreted as the stimulating effect due to presence of both actual thermal and Unruh thermal baths (this is inspired by an existing interpretation done earlier for the response function of an accelerating detector in a thermal bath \cite{Kolekar:2013hra,Kolekar:2013aka}). This structure of Feynman propagators implies that in presence of interaction the physical quantities in Rindler frame will  not only be affected by acceleration which is exactly like the same due to $\beta_0$, but also a stimulating effect must show its presence.

Among the two limits, $\beta_U\to\infty$ and $\beta_0\to\infty$, we observe that the later one is very interesting. 
For this, without applying analytic continuation, the Rindler propagators reduce to those in TFD with the identifications $\phi^R\equiv\phi$, $\phi^L\equiv\tilde{\phi}$ and $\beta_U \equiv \beta$. Thus these must be identical to those in CTP for $\sigma=\beta_U/2$. Although the result is known, but our approach is different from the earlier analysis \cite{Fulling1987135}. Previously this has been shown through defining Minkowski vacuum in terms of Rindler field states, whereas we show by calculating Feynman propagators. However there are advantages in our approach. We able to show more results of CTP through Rindler frame field theory. Application of analytic continuation provides us propagators which are compatible to two more specific values of $\sigma$ in CTP, namely $\sigma=0$ and $\sigma=\beta_U$. 
Therefore our investigation suggests a profound correspondence between the accelerated and thermal systems. Such an observation has various implications. These are as follows.
\begin{itemize}
\item Unlike TFD, in the two Rindler wedges we have two physical fields and therefore these are parts of our system. Moreover the thermal vacuum can be constructed from composite Fock space of these fields. Therefore such an analysis not only encodes the logistics of TFD, but also provides more physical picture in terms of the realization of both the fields. 
\item Similar to CTP, in our discussion we have a clear Keldysh contour for the Rindler time coordinate $\eta$. One part is the hyperbolic path in RRW, runs from $\eta=-\infty$ to $\eta=\infty$. The last part is the hyperbolic trajectory in LRW (runs from $\eta=\infty$ to $\eta=-\infty$) which is the mirror image of that in RRW. They are connected by a complex branch from $\eta=\infty$ to $\eta = \infty - i\beta_U/2$ for non-analytic continuation case. Therefore Rindler frame QFT also encodes the language of CTP. But here, unlike CTP, both the fields commute with each other.
\end{itemize}
In summary, the zero temperature ($\beta_0\to\infty$) field theory in Rindler frames, at least in the non-interacting level and for non-analytic continuation, can completely mimic the TFD formalism with added advantages like both the fields are now parts of the original system and can also encode features of CTP. Use of analytic continuation provides us freedom to go beyond TFD results as well as shows close proximity of CTP by invoking few more cases in CTP.

Let us recapitulate the main differences between the existing method \cite{Fulling1987135} and our present analysis towards these similarities among different formalisms. Although the thermal properties observed by an accelerated observer and the Minkowski vacuum can be expressed in terms of the right and left Rindler number states, thereby connection with TFD are well known, the correspondence between the field theory in the Rindler frame and the TFD formalism has not been adequately studied at the Feynman propagator level. Moreover, we consider an additional thermal bath of temperature $\beta_0$ to make things more generalised. The propagators in this generalised version have one non-thermal component, two thermal parts individually due to thermal bath and acceleration, and one cross-thermal part due to both thermal bath and acceleration. The incorporation of a real thermal bath is new and provides a completely distinct feature by introducing a cross-term in the propagators. At $\beta_0\to\infty$ limit, the Rindler propagators have a one-to-one correspondence with TFD propagators when one is not using the analytic continuation. Note that the TFD formalism is a particular case ($\sigma=\beta/2$) of the real-time formalism or CTP. However the introduction of analytic continuation in our approach shows new features, particularly we have been able to give a connection with CTP for two more values of $\sigma$. Any direct connection between CTP and the Rindler field theory is not well understood. Here, we furnish a possible close-time contour structure for the Rindler field theory, which is very similar to the CTP contour structure. 
Therefore we feel that this can be a deserving formalism to deal thermal theory of fields and also can be possible candidate to bridge between CTP and TFD.

We organize this paper in the following way. In Sec. \ref{sec:2}, a brief review on TFD and CTP is presented. Sec. \ref{Rindler} contains the main results regarding the thermal propagators in Rindler frame with respect to Minkowski vacuum. In this section we discuss the implications of our investigation and possible relation with TFD and CTP.  Finally, in Sec. \ref{Discussion}, we conclude.


\section{Brief review on the thermal field}\label{sec:2}
The ensemble average value of an observable $A$ for a system (denoted by Hamiltonian $H =  \sum_{\textbf{k}}\omega_{\textbf{k}} a_{\textbf{k}}^{\dagger}a_{\textbf{k}}$) in thermal equilibrium at inverse temperature $\beta$ is given by $\langle A_{}\rangle_{\beta}={\text{Tr}(e^{-\beta H_{}}A_{})}/{\text{Tr}(e^{-\beta H_{}})}$. In this formulation the thermal Green's function in field theory can be expressed as the sum of a infinite series for the Green's function corresponding to zero temperature field theory with the complexification of time. This is known as Matsubara \cite{Matsubara:1955ws} or the imaginary time formalism. On the other hand such an averaging process can be equivalently described by introducing a thermal vacuum state ($|0,\beta\rangle$) such that $\langle0,\beta|A_{}|0,\beta\rangle=\langle A_{}\rangle_{\beta}$. In order to construct the state, one introduces a fictitious system, which is identical to the original system, except it follows a tilde conjugation rule \cite{DasFTFT,10.1143/PTP.124.95}. This is known as thermofield double state (TFD) formalism. For the scalar field, this is given by \cite{DasFTFT}
\begin{equation}
|0,\beta\rangle=U(\theta)|0,\tilde{0}\rangle=e^{-\sum_{\textbf{k}}\theta_{\textbf{k}}(\beta)(\tilde{a}_{\textbf{k}}a_{\textbf{k}}-a_{\textbf{k}}^{\dagger}\tilde{a}_{\textbf{k}}^{\dagger})}|0,\tilde{0}\rangle\,
=\frac{1}{\sqrt{Z_{}(\beta)}}\sum_{\textbf{k},n_{\textbf{k}}}e^{-\beta n_{\textbf{k}}\omega_{\textbf{k}}/2}|n_{\textbf{k}},\,\tilde{n}_{\textbf{k}}\rangle\,~,
\label{B3}
\end{equation}
where $Z_{}(\beta)=\sum_{\textbf{k}}(1-e^{-\beta\omega_{\textbf{k}}})^{-1}$. In the above tilde denotes quantities for fictitious system whereas without tilde ones are for original system. $|n_{\textbf{k}},\,\tilde{n}_{\textbf{k}}\rangle\,$ is the Fock state for the composite system. The variable $\theta$ is function of $\beta$, and is given by $\tanh\theta_{\textbf{k}}(\beta)=e^{-\beta\omega_{\textbf{k}}/2}$.

In this formalism the thermal propagator is defined through $iG_{\beta}(x,y)=\langle0,\beta|\mathcal{T}[\Phi(x)\Phi^{T}(y)]|0,\beta\rangle$ where 
\begin{equation}
\Phi (x)=\begin{pmatrix}
\phi (x)\\\tilde{\phi}(x)
\end{pmatrix}
\label{B1}
\end{equation} 
is a doublet constructed by original field $\phi(x)$ and tilde field $\tilde{\phi}(x)$ configurations. $\mathcal{T}$ and $T$ denote time-ordering and transpose, respectively. The propagator is represented by a $2\times 2$ matrix. In momentum space this is given by 
\begin{equation}
\label{TFf1}
\begin{aligned}
 G_{\beta}(k)&=
  \begin{pmatrix}
\frac{1}{k^{2}-m^{2}+i\epsilon}&0\\0&-\frac{1}{k^{2}-m^{2}-i\epsilon}
\end{pmatrix}
-2i\pi\,n_{\beta}(|k^{0}|)\,\delta(k^{2}-m^{2})\begin{pmatrix}
1&e^{\beta|k^{0}|/2}\\e^{\beta|k^{0}|/2}&1
\end{pmatrix}\,,
 \end{aligned}\end{equation}
 where $n_{\beta}(|k^0|)=(e^{\beta\omega_{\textbf{k}}}-1)^{-1}$ is the bosonic number density. Here $k^2$ is the square of norm of the four-momentum, $m$ is the mass of the field and $|k^0| = \omega_{\textbf{k}} = (\textbf{k}^2 + m^2)^{1/2}$. The first and last diagonal elements are Feynman propagators for $\phi$ and $\tilde{\phi}$, respectively. The off-diagonal terms are ($\langle\phi\tilde{\phi}\rangle$ (top-right term) and $\langle\tilde{\phi}\phi\rangle$ (bottom-left term)) cross propagators between the original and the tilde fields.
 The thermal propagator has two contributions: the non-thermal propagator and the temperature dependent contributions. The thermal contribution contains bosonic number density and a Dirac delta function selecting only the on-shell modes. Due to the thermal contributions the off-diagonal terms are also non-vanishing.

\begin{figure}[h!]
    \centering
   {\includegraphics[width=0.49\textwidth]{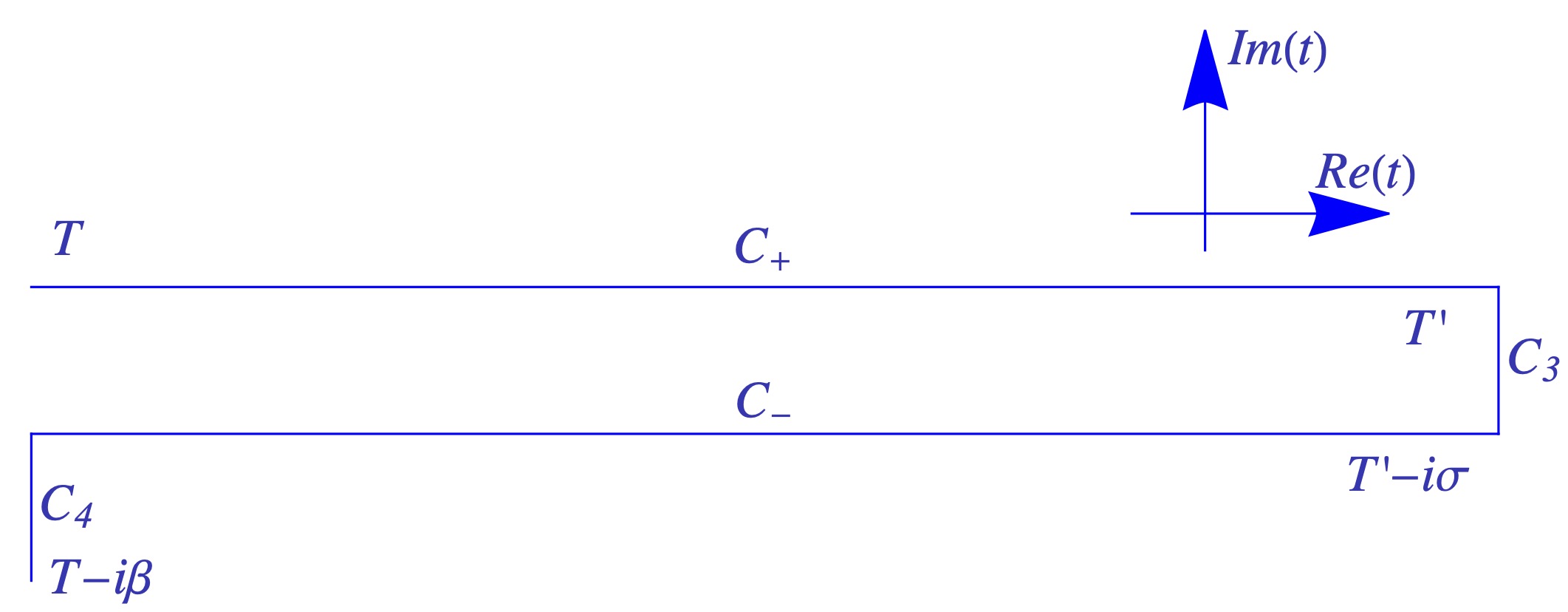}} 
   \caption{Closed time contour of Schwinger-Keldysh formalism with real branches $C_{+}$ and $C_{-}$, imaginary branches $C_{3}$ and $C_{4}$. Here one take the limits $T\to-\infty$ and $T'\to\infty$. }
   \label{fig:contour} 
\end{figure}

 Another approach to thermal field theory which is reminiscent to TFD is the closed time path (CTP) formalism. Such a formalism is based on a particular choice of Keldysh contour. For the choice, as shown in Fig. \ref{fig:contour}
 one introduces two-field configurations. The parts of the contour can be named as $C_{+}$, $C_{-}$, $C_{3}$ and $C_{4}$. On $C_{+}$ and $C_{-}$ branches time runs from $-\infty$ to $\infty$ and from $\infty$ to $-\infty$, respectively. In the imaginary time branches $C_{3}$ and $C_{4}$ time ranges are $(\infty$,$\infty-i\sigma$) and ($-\infty-i\sigma$, $-\infty-i\beta$), respectively with $0\leq \sigma\leq \beta$. We denote the field lying on $C_+$ branch as $\phi_+$ and the same on $C_-$ is labeled as $\phi_-$. In this formalism the propagator is represented by $2\times 2$ matrix as well and is given by
\begin{eqnarray}
\label{TF13}
G(k)=
\begin{pmatrix}
G_{++}&G_{+-}\\G_{-+}&G_{--} 
\end{pmatrix}
=
\begin{pmatrix}
\frac{1}{k^{2}-m^{2}+i\epsilon}&0\\0&-\frac{1}{k^{2}-m^{2}-i\epsilon}
\end{pmatrix}\nonumber
\\
-2i\pi\,\delta(k^{2}-m^{2})\begin{pmatrix}
n_{\beta}(\omega_{\textbf{k}})\,&e^{\sigma\omega_{\textbf{k}}}(n_{\beta}(\omega_{\textbf{k}})\, + \Theta(-\omega_\textbf{k}))\\e^{-\sigma\omega_{\textbf{k}}}(n_{\beta}(\omega_{\textbf{k}})\, + \Theta(\omega_\textbf{k}))&n_{\beta}(\omega_{\textbf{k}})\,
\end{pmatrix}\,.
\end{eqnarray}
Here, $\Theta$ is the Heaviside step function. The first-row elements are for $(t,t'\in C_{+})$ and $(t,\in C_{+};\,t'\in C_{-})$, respectively. While those in the second row are for $(t,\in C_{-};\,t'\in C_{+})$ and $(t,t'\in C_{-})$, respectively.

Although (\ref{TF13}) are (\ref{TFf1}) in general different, but for a specific choice $\sigma=\beta/2$ they coincide. In this particular case it appears that $\phi=\phi_+$ and $\tilde{\phi}$ gets a physical identity as $\phi_-$ and hence TFD and CTP might be equivalent for such a spacial situation. But in reality these are not physically equivalent. In TFD different fields commute whereas the commutation relation between two types of fields in CTP is different. Also in presence of interactions, these two formalisms behave differently.

So far we observed that TFD and CTP formalisms, although physically different, coincide at free theory level for a certain choice of $\sigma$. Moreover there are certain advantages or disadvantages in the respective formulations. For example, TFD is capable of describing the thermal-vacuum state; but deals with a fictitious field. Whereas as CTP dose suffer from the first issue while it is free of later one. Then naturally one might be wondering whether there is any other way of describing thermal field theory which can provide a middle root such that those issues can be tackled in a better way. We found that field theory in Rindler frame can be one such candidate. Let us now provide the motivation for such possibility and show how far we can proceed.

\section{Free field theory in Rindler frame}\label{Rindler}
In $(1+1)$-dimensions trajectory of a uniformly accelerated observer with uniform acceleration $a$ is given by the coordinate transform
\begin{equation}\label{TF30}
t = \frac{e^{a\xi}}{a} \sinh{a\eta};~~ x= 
\frac{e^{a\xi}}{a} \cosh{a\eta}\,,
\end{equation}
where $(t,x)$ are the Minkowski coordinates. And $(\eta,\,\xi)$ are the Rindler coordinates, which are extended from $-\infty$ to $\infty$. However these coordinate system covers only a single quadrant (where $|t|<x$) of the total Minkowski spacetime. This region is called the RRW. In a similar way one may get another Rindler wedge by the following transformation
\begin{equation}\label{TF31}
t =- \frac{e^{a\xi'}}{a} \sinh{a\eta'};~~ x= -
\frac{e^{a\xi'}}{a} \cosh{a\eta'}\,,
\end{equation}
where the coordinates covers the region bounded by $|t|<-x$. This region is known as the LRW. There are another two Rindler wedges, known as the future and past Rindler wedges (FRW and PRW). Since timelike trajectories can be formed only in RRW and LRW, it is then natural to consider two observers which can be thought of two systems. Under these frames one can the introduce two species of fields. Therefore such two accelerated frames can describe two physically existing fields -- one is lying in RRW and another is in LRW. Also note that one set of coordinates go to other set (defined in (\ref{TF30}) and (\ref{TF31})) through the analytic continuation 
\begin{equation}
\eta \rightarrow \eta'\pm i\frac{\pi}{a}~.
\label{analytic}
\end{equation} 
We will use this later in our analysis.

The trajectories of an accelerated observers are provided by a hyperbola as shown in Fig. \ref{fig:contour2}. Coordinate transformation of the both regions give the same line element as
\begin{equation}\label{TF32}
 ds^2 = e^{2a\xi}\left[d\eta^2-d\xi^2\right]\,,
\end{equation}
which is conformally flat and one obtain the massive Klein-Gordon (KG) modes in the right and left Rindler wedges as \cite{book:Birrell, book:carroll}
\begin{equation}\label{TF33}\begin{aligned}
 ^{R}u_{k} &=\frac{1}{\sqrt{4\pi\omega_{\textbf{k}}}} 
e^{i\textbf{k}\xi-i\omega_{\textbf{k}}\eta}~~~\textup{in~RRW}\\ 
~&= 0~~~~~~~~~~~~~~~~~~~~\textup{in~LRW}\\
 ^{L}u_{k} &= \frac{1}{\sqrt{4\pi\omega_{\textbf{k}}}} 
e^{i\textbf{k}\xi+i\omega_{\textbf{k}}\eta}~~~\textup{in~LRW}\\
~&= 0~~~~~~~~~~~~~~~~~~~~\textup{in~RRW}.
\end{aligned}\end{equation}
These sets of modes are complete in either RRW or LRW, not in the full Minkowski space. However they together form complete basis for the whole spacetime. Using them one can find two physical entities of field configuration corresponding to two Rindler observers:
\begin{equation}\label{FT34}\begin{aligned}
 \phi^{R}(x) &=\sum_{\textbf{k}=-\infty}^{\infty} 
\left[b^{R}_{\textbf{k}}~ ^{R}u_{\textbf{k}} + 
b^{R^{\dagger}}_{\textbf{k}}~ ^{R}u_{\textbf{k}}^{*}\right]\,,\\
\phi^{L}(x) &= \sum_{\textbf{k}=-\infty}^{\infty} 
\left[b^{L}_{\textbf{k}}~ ^{L}u_{\textbf{k}} + 
b^{L^{\dagger}}_{\textbf{k}}~ ^{L}u_{\textbf{k}}^{*}\right] \,.
\end{aligned}\end{equation}
Superscript $L$ and $R$ correspond to the left and the right Rindler wedges respectively, and the annihilation operators $b_{\textbf{k}}^{R,L}$ annihilate the Rindler vacuum $|0_{\mathcal{R}}\rangle$.
To make life easy it will be better to work with Unruh operators $(d^{1}_{k}, d^{2}_{k})$ which annihilate the Minkowski Vacuum $(d^{1}_{\textbf{k}}|0_{M}\rangle= d^{2}_{\textbf{k}}|0_{M}\rangle=0)$, provided by Unruh \cite{Unruh:1976db} in $1976$. 
Then the above becomes \cite{book:Birrell}
\begin{equation}
\label{TF36}
\begin{aligned}
 \phi^{R}(x) &= \sum_{\textbf{k}=-\infty}^{\infty} 
\frac{1}{\sqrt{2\sinh{\frac{\pi\omega_{\textbf{k}}}{a}}}} 
\left[d^{1}_{\textbf{k}}~e^{\frac{\pi\omega}{2a}}~ ^{R}u_{\textbf{k}} 
+ d^{2}_{\textbf{k}}~e^{-\frac{\pi\omega}{2a}}~ ^{R}u^{*}_{-\textbf{k}} \right] + 
h.c.\,,\nonumber\\
\phi^{L}(x) &= \sum_{\textbf{k}=-\infty}^{\infty} 
\frac{1}{\sqrt{2\sinh{\frac{\pi\omega_{\textbf{k}}}{a}}}} 
\left[d^{1}_{\textbf{k}} 
e^{-\frac{\pi\omega_{\textbf{k}}}{2a}}~ ^{L}u^{*}_{-\textbf{k}}
+ d^{2}_{\textbf{k}} 
e^{\frac{\pi\omega_{\textbf{k}}}{2a}}~ ^{L}u_{\textbf{k}} \right] + 
h.c.\,.
\end{aligned}
\end{equation}
In the above $h.c.$ refers to hermitian conjugate. 

\subsection{Without using analytic continuation}

In this discussion for generality, we consider the fields to be in thermal equilibrium with an external thermal bath whose inverse temperature is $\beta_0$. Then defining the Wightman function with respect to Minkowski vacuum as $G_{W_{ij}} (x_1,x_2) = \langle0_{M}|e^{-\beta_0H}\phi^i(\eta_1,\xi_1)\phi^j(\eta_2,\xi_2)|0_{M}\rangle/Z(\beta_{0})$ under the concept of ensemble average of the field operators, we have four propagators. We name the propagators as $G_{W_{RR}},\,G_{W_{RL}},\,G_{W_{LR}}$ and $G_{W_{LL}}$. We also consider equal magnitudes of the acceleration for both observers. Considering the Hamiltonian $H$ corresponding to the Unruh modes we obtain the following Wightman functions \cite{Barman:2021bbw} 
\begin{equation}\label{TF37}\begin{aligned}
		G_{W_{RR}}\left(\Delta\xi,\Delta\eta\right)= &\int_{-\infty}^{\infty} 
		\frac{d\textbf{k}~n_{\beta_{0}}(\omega_{\textbf{k}}) }{8\pi\omega_{\textbf{k}}\sinh{\frac{\pi\omega_{\textbf{k}}}{a}}}
		\left[e^{\beta_{0}\omega_{\textbf{k}}}(\omega)\left\{e^{i\textbf{k}\Delta\xi-i\omega_{\textbf{k}}\Delta\eta} ~e^{\frac{\pi\omega_{\textbf{k}}}{a}}+
		e^{i\textbf{k}\Delta\xi+i\omega_{\textbf{k}}\Delta\eta}~
		e^{-\frac{\pi\omega_{\textbf{k}}}{a}}\right\}\right.\\&
		\left.~~~~~~~~~~~~~~~~~~~~~~~+
		\left\{e^{-i\textbf{k}\Delta\xi+i\omega_{\textbf{k}}\Delta\eta} 
		~e^{\frac{\pi\omega_{\textbf{k}}}{a}} +~
		e^{-i\textbf{k}\Delta\xi-i\omega_{\textbf{k}}\Delta\eta}~
		e^{-\frac{\pi\omega_{\textbf{k}}}{a}}\right\}\right]\,;
		\\      
			G_{W_{LL}}\left(\Delta\xi,\Delta\eta\right)= &\int_{-\infty}^{\infty} 
		\frac{d\textbf{k}~n_{\beta_{0}}(\omega_{\textbf{k}}) }{8\pi\omega_{\textbf{k}}\sinh{\frac{\pi\omega_{\textbf{k}}}{a}}}
		\left[e^{\beta_{0}\omega_{\textbf{k}}}\left\{e^{i\textbf{k}\Delta\xi+i\omega_{\textbf{k}}\Delta\eta} ~e^{\frac{\pi\omega_{\textbf{k}}}{a}}+
		e^{i\textbf{k}\Delta\xi-i\omega_{\textbf{k}}\Delta\eta}~
		e^{-\frac{\pi\omega_{\textbf{k}}}{a}}\right\}\right.\\&
		\left.~~~~~~~~~~~~~~~~~~~~~~~+
		\left\{e^{-i\textbf{k}\Delta\xi-i\omega_{\textbf{k}}\Delta\eta} 
		~e^{\frac{\pi\omega_{\textbf{k}}}{a}} +~
		e^{-i\textbf{k}\Delta\xi+i\omega_{\textbf{k}}\Delta\eta}~
		e^{-\frac{\pi\omega_{\textbf{k}}}{a}}\right\}\right]\,;
		\\	
				G_{W_{LR}}\left(\Delta\xi,\Delta\eta\right)=& \int_{-\infty}^{\infty} 
		\frac{d\textbf{k}~n_{\beta_{0}}(\omega_{\textbf{k}}) }{8\pi\omega_{\textbf{k}}\sinh{\frac{\pi\omega_{\textbf{k}}}{a}}}
		\left[e^{\beta_{0}\omega_{\textbf{k}}}\left\{e^{i\textbf{k}\Delta\xi-i\omega_{\textbf{k}}\Delta\eta}+
		e^{i\textbf{k}\Delta\xi+i\omega_{\textbf{k}}\Delta\eta}\right\}\right.\\&
		\left.~~~~~~~~~~~~~~~~~~~~~~~+
		\left\{e^{-i\textbf{k}\Delta\xi+i\omega_{\textbf{k}}\Delta\eta} +~
		e^{-i\textbf{k}\Delta\xi-i\omega_{\textbf{k}}\Delta\eta}~\right\}\right]\,;
		\\	
			G_{W_{RL}}\left(\Delta\xi,\Delta\eta\right)=& \int_{-\infty}^{\infty} 
		\frac{d\textbf{k}~n_{\beta_{0}}(\omega_{\textbf{k}})}{8\pi\omega_{\textbf{k}}\sinh{\frac{\pi\omega_{\textbf{k}}}{a}}}
		\left[e^{\beta_{0}\omega_{\textbf{k}}}\left\{e^{-i\textbf{k}\Delta\xi+i\omega_{\textbf{k}}\Delta\eta}+
		e^{-i\textbf{k}\Delta\xi - i\omega_{\textbf{k}}\Delta\eta}\right\}\right.\\&
		\left.~~~~~~~~~~~~~~~~~~~~~~~+
		\left\{e^{i\textbf{k}\Delta\xi-i\omega_{\textbf{k}}\Delta\eta} +~
		e^{i\textbf{k}\Delta\xi+i\omega_{\textbf{k}}\Delta\eta}~\right\}\right]\,.
		\end{aligned}\end{equation}

Having these positive frequency Wightman functions, we can now define the respective Feynman-like propagators through the relation
\begin{equation}\label{TF38a}\begin{aligned}
iG_{{ij}}\left(\Delta\xi,\Delta\eta\right)= &\Theta(\Delta\eta)G_{W_{ij}}\left(\Delta\xi,\Delta\eta\right)+\Theta(-\Delta\eta)G_{W_{ij}}\left(\Delta\xi,-\Delta\eta\right) \,,
\end{aligned}\end{equation}
where the subscripts are $i,\,j=R,\,L$. In $4$-momentum space these are given by
\begin{equation}\label{TF38}\begin{aligned}
iG_{{RR}}(
{k})&=
\int_{-\infty}^{\infty}d\Delta\eta\int_{-\infty}^{\infty}d\Delta\xi\,iG_{{RR}}\left(\Delta\xi,\Delta\eta\right)\,e^{-i{\textbf{k}}\Delta\xi+i{\omega}\Delta\eta}\\&=
i\,
\left\{\frac{1}{({\omega}^{2}-\omega_{{\textbf{k}}}^{2}+i\epsilon)}  -2i\pi\,n_{\beta_{0}}\,\delta({\omega}^{2}-\omega_{{\textbf{k}}}^{2})
-2i\pi\,n_{a}\,\delta({\omega}^{2}-\omega_{{\textbf{k}}}^{2})
-4i\pi\, n_{\beta_{0}}\,n_{a}\,\delta({\omega}^{2}-\omega_{{\textbf{k}}}^{2})\right\}\\&=
i\,
\left\{\frac{1}{({k}^{2}-m^{2}+i\epsilon)}  -2i\pi\,n_{\beta_{0}}\,\delta({k}^{2}-m^{2})
-2i\pi\,n_{a}\,\delta({k}^{2}-m^{2})
-4i\pi\, n_{\beta_{0}}\,n_{a}\,\delta({k}^{2}-m^{2})\right\}\,,
\end{aligned}\end{equation}
\begin{equation}\label{TF39}\begin{aligned}
iG_{{LL}}({k})&=
\int_{-\infty}^{\infty}d\Delta\eta\int_{-\infty}^{\infty}d\Delta\xi\,iG_{{LL}}\left(\Delta\xi,\Delta\eta\right)\,e^{-i{\textbf{k}}\Delta\xi+i{\omega}\Delta\eta}\\&=i\,
\left\{-\frac{1}{({\omega}^{2}-\omega_{{\textbf{k}}}^{2}-i\epsilon)}  -2i\pi\,n_{\beta_{0}}\,\delta({\omega}^{2}-\omega_{{\textbf{k}}}^{2})
-2i\pi\,n_{a}\,\delta({\omega}^{2}-\omega_{{\textbf{k}}}^{2})
-4i\pi\, n_{\beta_{0}}\,n_{a}\,\delta({\omega}^{2}-\omega_{{\textbf{k}}}^{2})\right\}\\&=i\,
\left\{-\frac{1}{({k}^{2}-m^{2}-i\epsilon)}  -2i\pi\,n_{\beta_{0}}\,\delta({k}^{2}-m^{2})
-2i\pi\,n_{a}\,\delta({k}^{2}-m^{2})
-4i\pi\, n_{\beta_{0}}\,n_{a}\,\delta({k}^{2}-m^{2})\right\}\,.
\end{aligned}\end{equation}
\begin{equation}\label{TF40}\begin{aligned}
iG_{{RL}}(
{k})&=
\int_{-\infty}^{\infty}d\Delta\eta\int_{-\infty}^{\infty}d\Delta\xi\,iG_{{RL}}\left(\Delta\xi,\Delta\eta\right)\,e^{-i{\textbf{k}}\Delta\xi+i{\omega}\Delta\eta}\\ 
&=
i\,e^{\frac{\pi\omega_{{\textbf{k}}}}{a}}\,n_{a}\,
	\left(2\,n_{\beta_{0}}+1\right)\left[-2i\pi\delta({\omega}^{2}-\omega_{{\textbf{k}}}^{2})\right]\\ 
	&=
i\,e^{\frac{\pi\omega_{{\textbf{k}}}}{a}}\,n_{a}\,
	\left(2\,n_{\beta_{0}}+1\right)\left[-2i\pi\delta({k}^{2}-m^{2})\right]\, = iG_{{LR}}(
{k}).
\end{aligned}\end{equation}
In the above we define ${k}=({\omega},\,{\textbf{k}})$, $\omega_{{\textbf{k}}}^{2}={\textbf{k}}^{2}+m^{2}$, $n_{\beta_0} = (e^{\beta_0\omega_{\bf{k}}}-1)^{-1}$ and $n_{a} = (e^{\frac{2\pi\omega_{\bf{k}}}{a}}-1)^{-1}$. Explicit steps to obtain the above forms are given in Appendix \ref{A}. Note that in calculating these Feynman propagators we did not use analytic continuation relations (\ref{analytic}) among the coordinates in Wightman functions.
Here both $iG_{{RR}}({k})$ and $iG_{{LL}}({k})$ have four terms. The first term is independent of temperature and acceleration of the detectors. This represents the propagator for inertial motion of the observers in the zero-temperature field theory. The second term has only temperature dependent. The first and second terms together represent the propagator for inertial observers in finite temperature. The third term has only acceleration dependent, shows the contribution for accelerated motion of the detectors in the zero-temperate field theory. However the last term contains both temperature and motion dependence. Moreover, these are symmetric under the exchange of $\beta_0 \leftrightarrow \beta_U = 2\pi/a$, where $\beta_U$ is the inverse Unruh temperature. This is the reflection of the well known fact that the field theory in indivitual Rindler frame mimics thermal field theory. On the other hand other two are not in similar structure. Presence of cross terms in the propagators implies that the calculation of the physical quantities in interaction scenario with respect to Rindler frame not only contributed by acceleration effect which is identical to thermal field theory in Minkowski frame, but also by a stimulating effect in which $\beta_0$ and $\beta_U$ are stimulating each other.
Let us now consider two important limits -- one is $\beta_0\to\infty$ and other one is $\beta_U\to\infty$.

For the first limit (corresponds to zero temperature field theory in Rindler frame) the above reduces to
\begin{equation}\label{TF381}\begin{aligned}
G_{{RR}}(
{k})=
\frac{1}{({k}^{2}-m^{2}+i\epsilon)} -2i\pi\,n_{a}\,\delta({k}^{2}-m^{2})
\,,
\end{aligned}\end{equation}
\begin{equation}\label{TF391}\begin{aligned}
G_{{LL}}(
{k})=-
\frac{1}{({k}^{2}-m^{2}-i\epsilon)} -2i\pi\,n_{a}\,\delta({k}^{2}-m^{2})
\,,
\end{aligned}\end{equation}
\begin{equation}\label{TF401}\begin{aligned}
G_{{RL}}(
{k})=
e^{\frac{\pi\omega_{{\textbf{k}}}}{a}}\,n_{a}\,
	\left[-2i\pi\delta({k}^{2}-m^{2})\right]\, = G_{{LR}}(
{k})\,.
\end{aligned}\end{equation}
Arranging these in the following matrix form:
\begin{equation}\label{B2}
G=\begin{pmatrix}
G_{RR}&G_{RL}\\G_{LR}&G_{LL}
\end{pmatrix}\,.
\end{equation} \begin{figure}[h!]
    \centering
   {\includegraphics[width=0.49\textwidth]{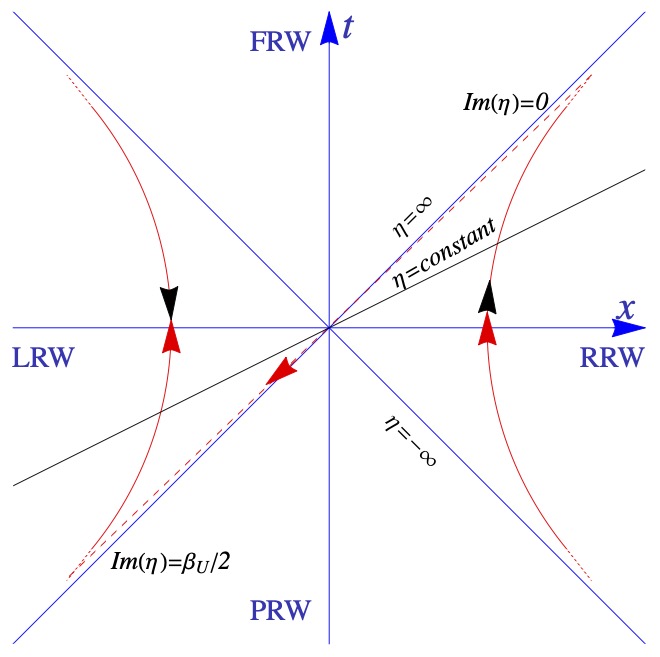}} 
   \caption{Trajectory of an accelerated detector in RRW and LRW are shown as red hyperbolas. The black arrows represent direction of the Rindler time propagation. The Rindler time ($\eta$) in RRW is increasing upward. In LRW, it is decreasing upward. The dashed red line represents a complex Rindler time propagation (consider it as an imaginary line in complex $\eta$-plane), connecting two Rindler time branches in RRW and LRW. The red arrows represent the direction of time ($\eta$) propagation in the closed contour.}
   \label{fig:contour2} 
\end{figure}
we observe that this is identical to that of TFD (see Eq. (\ref{TFf1})) with the identification $\beta_U = \beta$. Thus the propagators of two accelerated observers in right and left Rindler wedges in zero-temperature field theory have same structure as the thermal propagators obtained through TFD formalism. Moreover since both the fields have real existence, it has a quite resemblance with the CTP. The schematic Schwinger-Keldysh path in this case shown in Fig. \ref{fig:contour2} . In this case the transition from one part of path to the other part is done by a fixed value $\sigma=\beta_U/2$ and hence, unlike CTP formalism, can not be any arbitrary value between $0$ and $\beta_U$. In this regard, it may be pointed out that exactly the same value was taken for time to connect the two Rindler wedges in discussing Unruh effect through path integral approach \cite{Padmanabhan:2016xjk}. The thermal vacuum is then given by (\ref{B3}) with  the Fock basis is identified as $|n_{\textbf{k}},\,\tilde{n}_{\textbf{k}}\rangle = |n^R_{\textbf{k}},\,n^L_{\textbf{k}}\rangle$ which are in Rindler frames and $|0,\beta\rangle = |0_{M}\rangle$. 

For the other limit $\beta_U\to \infty$ (corresponds to inertial motions of the observers), the matrix (\ref{B2}) becomes diagonal. $G_{RR}$ takes the form of usual thermal Feynman propagator and $G_{LL}$ becomes corresponding anti-Feynman propagator.

\subsection{Incorporating analytic continuation: correspondence with other $\sigma$-values of CTP}
As we already discussed earlier that in CTP formalism, one can take an arbitrary fixed complex separation ($i\sigma$) between the parts $C_{+}$ and $C_{-}$. The green functions for arbitrary value of $\sigma$ are given in Eq. (\ref{TF13}). Among all possible $\sigma$ values, $\sigma=\beta/2$ corresponds for TFD. In Rindler field theory, as shown in the above the same is obtained without using the analytic continuation (\ref{analytic}) in the Wightman functions and this corresponds to $\sigma = \beta_U/2 = \pi/a$. Now we are going to use the analytic continuation (\ref{analytic}) on one of the time ordinates among $\eta$ and $\eta'$ in Wightman functions (\ref{TF37}) before calculating the Feynman propagators in momentum space. We will use on $\eta'$; i.e. $\eta'\to \eta \pm i\pi/a$. So we have $\eta_1\to\eta_1$ and $\eta'_2\to \eta_2\pm i\pi/a$. Under this it is easy to check that $G_{RR}(k)$ and $G_{LL}(k)$ will not be affected and they are given by (\ref{TF38}) and (\ref{TF39}), respectively. 

However, the cross green's functions will be affected.
In this case one has to replace $\Delta \eta \to \Delta\eta - i\epsilon$, with two values of $\epsilon$ as $\epsilon =\pm\pi/a$.
Then we find
\begin{equation}\label{TF403}\begin{aligned}
iG_{{RL}}({k})&=
\int_{-\infty}^{\infty}d\Delta\eta\int_{-\infty}^{\infty}d\Delta\xi\,iG_{{RL}}\left(\Delta\xi,\Delta\eta-i\epsilon\right)\,e^{-i{\textbf{k}}\Delta\xi+i{\omega}\Delta\eta}\\ &=
i\,e^{\frac{\pi\omega_{{\textbf{k}}}}{a}-\omega\epsilon}\,n_{a}\,
	\left(2\,n_{\beta_{0}}+1\right)\left[-2i\pi\delta({\omega}^{2}-\omega_{{\textbf{k}}}^{2})\right]\\ &=
i\,e^{\frac{\pi\omega_{{\textbf{k}}}}{a}-\omega\epsilon}\,n_{a}\,
	\left(2\,n_{\beta_{0}}+1\right)\left[-2i\pi\delta({k}^{2}-m^{2})\right]\,;\\
iG_{{LR}}({k})&=
\int_{-\infty}^{\infty}d\Delta\eta\int_{-\infty}^{\infty}d\Delta\xi\,iG_{{RL}}\left(\Delta\xi,\Delta\eta+i\epsilon\right)\,e^{-i{\textbf{k}}\Delta\xi+i{\omega}\Delta\eta}\\ &=
i\,e^{\frac{\pi\omega_{{\textbf{k}}}}{a}+\omega\epsilon}\,n_{a}\,
	\left(2\,n_{\beta_{0}}+1\right)\left[-2i\pi\delta({\omega}^{2}-\omega_{{\textbf{k}}}^{2})\right]\\ &=
i\,e^{\frac{\pi\omega_{{\textbf{k}}}}{a}+\omega\epsilon}\,n_{a}\,
	\left(2\,n_{\beta_{0}}+1\right)\left[-2i\pi\delta({k}^{2}-m^{2})\right]\,.
\end{aligned}\end{equation}
In this case also $G_{RR}$ and $G_{LL}$ are symmetric under the exchange between $\beta_0$ and $\beta_U$, while the above two are not.
Here again these green functions have two parts. One part only has acceleration dependence. Another part has both acceleration and temperature dependence. 

By setting the external temperature to $zero$ ($\beta_{0}\to\infty$ limit), these green functions will reduced to 
\begin{equation}\label{TF404}\begin{aligned}
iG_{{RL}}({k})&=
i\,e^{\frac{\pi\omega_{{\textbf{k}}}}{a}-\omega\epsilon}\,n_{a}\,
	\left[-2i\pi\delta({k}^{2}-m^{2})\right]\,;\\
iG_{{LR}}({k})&=
i\,e^{\frac{\pi\omega_{{\textbf{k}}}}{a}+\omega\epsilon}\,n_{a}\,
	\left[-2i\pi\delta({k}^{2}-m^{2})\right]\,.
\end{aligned}\end{equation}
We have two choices for $\epsilon$. Let us now discuss whether these two choices have any connection to CTP.
\vskip 1mm
\noindent
{\underline{Case I. $\epsilon=\pi/a$}}:
For this particular choice, (\ref{TF404}) reduces to
\begin{equation}\label{TF404-1}\begin{aligned}
iG_{{LR}}({k})&=
2\pi\,(n_{a}(|\omega|)+\Theta(\omega))\,
	\delta({k}^{2}-m^{2})\,,\\
iG_{{RL}}({k})&=
2\pi\,(n_{a}(|\omega|)+\Theta(-\omega))\,
	\delta({k}^{2}-m^{2})\,~.
\end{aligned}\end{equation}
Note that these are $iG_{-+}$ and $iG_{+-}$, respectively for $\sigma=0$ of CTP, given in (\ref{TF13}) with the identification $\beta = \beta_U = 2\pi/a$. 
\vskip 1mm
\noindent
{\underline{Case II. $\epsilon= - \pi/a$}}:
In this case (\ref{TF404}) yields
\begin{equation}\label{TF404-2}\begin{aligned}
iG_{{LR}}({k})&=
2\pi\,(n_{a}(|\omega|)+\theta(-\omega))\,
	\delta({k}^{2}-m^{2})\,,\\
iG_{{RL}}({k})&=
2\pi\,(n_{a}(|\omega|)+\theta(\omega))\,
	\delta({k}^{2}-m^{2})\,~,
\end{aligned}\end{equation}
which are $iG_{-+}$ and $iG_{+-}$, respectively for $\sigma=2\pi/a$ of CTP, given in (\ref{TF13}).
Note appearance of the parameter $\omega$, which comes through the Fourier transform and hence it can take both positive and negative values. In this calculation it takes the values $\omega_{\textbf{k}}$ or  $-\omega_{\textbf{k}}$, where $\omega_{\textbf{k}}>0$. 
 

In this section we observed that the non-thermal QFT on Rindler frame with no-analytic continuation provides TFD description, and hence $\sigma=0$ case of CTP. However we have a new finding that two other specific cases of CTP can be mimicked through QFT in Rindler frame by introducing analytic continuation.   
Therefore it seems that the field theory in Rindler frame, at least in absence of interaction, carries all necessary information of TFD and at least few properties of CTP. Hence we feel that Rindler frame analysis can be considered as a possible candidate for dealing thermal theory of fields. Moreover, it may act as a possible bridge between these two existing formalisms and thereby may provide a unified picture between them. However here, like in TFD but unlike in CTP, $\phi^R$ and $\phi^L$ commute.
\section{Conclusions}\label{Discussion}
Dealing with the thermal properties of a system through TFD and CTP formalisms has various applications in different branches of physics. Although such formulations are very robust, as mentioned earlier, these have their own ``advantages'' and ``disadvantages''. In general, they are different by construction and carry different information. However, at the propagator level, two formalisms match each other for a particular choice of complex path in CTP (denoted earlier as $\sigma=\beta/2$). Here, we tried illuminating some of the missing links in the respective formalisms and searched for a possible bridge between them. We found that the zero-temperature QFT in Rindler frames may be a possible candidate to provide a parallel formalism for the thermal description of fields. It may also help build a bridge between these two formulations.

We considered two accelerated observers in RRW and LRW with the same constant acceleration. For generality, the quantum fields are taken to be thermal with inverse temperature $\beta_0$. The possible four combinations for Feynman-like propagators have been calculated using the thermal ensemble averaging process. It has been observed that both $G_{RR}$ and $G_{LL}$ have two contributions due to the background and Unruh-thermal baths along with the usual non-thermal contribution. Additionally, there is a cross-term due to the simultaneous presence of real and Unruh thermal baths.
Interestingly, they are symmetric under the exchange of $\beta_0\leftrightarrow\beta_U$. The other two (cross) propagators do not have identical exchange symmetry. For these propagators, there are two terms; one term only depends on acceleration, and another on both acceleration and real thermal bath. Such a result has a strong implication for various physical processes in an accelerated frame. Particularly, the presence of cross-term will show stimulating effects in various field theoretic calculations concerning the Rindler observer.

Here we discussed two limiting cases: $\beta_U\to\infty$ and $\beta_0\to\infty$. The second one is of particular interest to us. In this limit, without incorporating analytic continuation between the RRW and LRW coordinates, we have shown that the Rindler propagators are equivalent to the TFD propagators with identification $\beta=\beta_{U}=2\pi/a$. The fields in the right and left Rindler wedges ($\phi_{R}$ and $\phi_{L}$) play the role of the real and fictitious fields of the TFD system ($\phi$ and $\tilde{\phi}$), respectively. Moreover, the Minkowski vacuum seen from the Rindler frame is equivalent to the thermal vacuum of TFD. Thus, a thermal field in the Minkowski vacuum observed in the Minkowski frame is equivalent to the non-thermal field in the Minkowski vacuum observed in the Rindler frame. However, the later one has added advantages. Here, unlike in TFD, both fields are the parts of the original system. 
In fact, it goes beyond that. Particularly for non-analytic continuation case, our formulation leads to CTP where  the Keldysh contour is identified as shown in Fig. \ref{fig:contour2}, hence one has the choice $\sigma=\beta_U/2$. In both Rindler wedges, coordinate times are increasing in their frames. However, the coordinate time in one wedge runs in the opposite direction with respect to the other wedge, similar to the CTP contour. 
Although this is more or less known in the literature, we have few new observations in our approach. It has been shown that the incorporation of analytic continuation yields two more specific situations of CTP, namely $\sigma=0$ and $\sigma = \beta = \beta_U$, which are the two extreme values of the $\sigma$-parameter. This observation was missing in the earlier analysis. This is important as the present investigation and findings may indicate that the generalised version of CTP formalism may have a close relation with QFT in the Rindler frame. Note that, in the CTP formalism, $\sigma$ can take values from $0$ to $\beta$. Out of these $\sigma$ values, $\sigma=0$ corresponds to the so-called original Schwinger-Keldysh formalism with temperature $\beta\,$ and $\sigma=\beta/2$ corresponds to the TFD formalism. These are the most critical cases of CTP formalism. 
Hence, in summary, the Rindler frame formulation of quantum field theory at zero temperature corresponds to TFD, but the fields capture a physical existence here. Furthermore, as it mimics a few particular cases of CTP, this may be considered a viable theory of thermal fields, which may illuminate the unknown bridge between TFD and CTP formalisms. In this sense, at least at the free theory level, the present prescription can be considered as a ``hybridisation'' of TFD and CTP. 

So far, these observations provide few suggestive implications. In order to justify them concretely, we need further investigation. In CTP, $\sigma$ is entirely arbitrary and runs from $0$ to $\beta$. Here, we showed that Rindler frame formulation corresponds to three particular choices of $\sigma$. We feel that this is due to the fact that the same acceleration is considered for the two frames. It would be interesting to investigate by considering different accelerations of RRW and LRW observers and check whether such a situation, with continuously varying one of the accelerations, can mimic all possible choices of $\sigma$.
Furthermore, given the Feynman propagators at a finite temperature, one can study the various interacting theories, particularly the physical processes, in Rindler frames. In the current analysis, we considered that the fields are in equilibrium with a background thermal bath. This generalisation can be useful in certain physical situations, $i.e.$, in the cosmological inflation phase, where the hot and dense universe expands with some acceleration.  The earlier vacuum can be seen as a thermal vacuum in the accelerated frame. Due to the background thermal bath, there will be two additional contributions: one for the effect of this thermal bath and another for the stimulating effect of both thermal bath and acceleration. The actual tree-level results may differ from ours since the acceleration rate of the cosmological inflation phase is not constant. In studying hot collapsing neutron stars, one can build similar approximate models with some acceleration for collapsing.  
Also, note that the present analysis has been done in $(1+1)$-spacetime dimensions. It would be interesting to extend the present analysis in $(3+1)$-spacetime and look for the implications. We leave these directions for future studies.

\vskip 4mm
\noindent
{\bf Acknowledgments:}
DB would like to acknowledge Ministry of Education, Government of India for providing financial support for his research via the Prime Minister's Research Fellows (PMRF) May 2021 scheme.


\begin{appendix}
\begin{center}
\section*{Appendix}
\end{center}

\section{Feynman propagators in momentum representation}\label{A}

To evaluate the Fourier transformation (FT) of the propagators given in Eq. (\ref{TF38a}) and (\ref{TF37}) one calculates
\begin{equation}\label{TFA1}
\int_{-\infty}^{\infty}d\Delta\eta\int_{-\infty}^{\infty}d\Delta\xi\,iG_{{ij}}\left(\Delta\xi,\Delta\eta\right)\,e^{-i\bar{\textbf{k}}\Delta\xi+i\bar{\omega}\Delta\eta}~.
\end{equation}
The variables $\bar{\textbf{k}}$ and $\bar{\omega}$ are arbitrary 2-vector i.e., $\bar{k}^{a}=(\bar{\omega},\,\bar{\textbf{k}})$. We often need the following results to evaluate the above:
\begin{equation}\label{TFa2}\begin{aligned} 
&\int_{-\infty}^{\infty}d\Delta\xi\,e^{-i\bar{\textbf{k}}\Delta\xi}e^{\pm i{\textbf{k}}\Delta\xi}=2\pi\delta(\bar{\textbf{k}}\pm \textbf{k})~;\\&
\int_{-\infty}^{\infty}d\Delta\eta\,\,\theta(\pm\Delta\eta)\,e^{i\bar{\omega}\Delta\eta}\,e^{i(\text{sgn}(\omega)\omega\pm i\epsilon)\Delta\eta}=\frac{\pm i}{\bar{\omega}+\text{sgn}(\omega)\omega\pm i\epsilon}~,~~~~~~~(\textrm{with}~~~ \epsilon>0)
\end{aligned}\end{equation}

{\underline{$G_{RR}$:}} -- 
To evaluate the FT of  $iG_{F_{RR}}\left(\Delta\xi,\Delta\eta\right)$, we use the expression of $G_{F_{RR}}\left(\Delta\xi,\Delta\eta\right)$ provided in Eq. (\ref{TF37}), (\ref{TF38a}). Then, the $\Delta\eta$ and $\Delta\xi$-integrations can be evaluated using the identities in Eq. (\ref{TFa2}). Then one obtains
\begin{equation}\label{TF}\begin{aligned}
iG_{RR}(\bar{k})&=\int_{-\infty}^{\infty}d\Delta\eta\int_{-\infty}^{\infty}d\Delta\xi\,iG_{{RR}}\left(\Delta\xi,\Delta\eta\right)\,e^{-i\bar{\textbf{k}}\Delta\xi+i\bar{\omega}\Delta\eta}\\&=
 i\int_{-\infty}^{\infty} \frac{d\textbf{k}~n_{\beta_{0}}(\omega_{\textbf{k}})}{4\omega_{\textbf{k}}\sinh{\frac{\pi\omega_{\textbf{k}}}{a}}}
 \\&
	\left\{e^{\beta_{0}\omega_{\textbf{k}}}\left[\left( \frac{\delta(\textbf{k}-\bar{\textbf{k}})}{(\bar{\omega}-\omega_{\textbf{k}}+i\epsilon)}-\frac{\delta(\textbf{k}-\bar{\textbf{k}})}{(\bar{\omega}+\omega_{\textbf{k}}-i\epsilon)}\right)\, e^{\frac{\pi\omega_{\textbf{k}}}{a}} +
	\left(\frac{\delta(\textbf{k}-\bar{\textbf{k}})}{(\bar{\omega}+\omega_{\textbf{k}}+i\epsilon)}-\frac{\delta(\textbf{k}-\bar{\textbf{k}})}{(\bar{\omega}-\omega_{\textbf{k}}-i\epsilon)}\right)
	e^{-\frac{\pi\omega_{\textbf{k}}}{a}}\right]\right.\\&+
	\left.\left[\left( \frac{\delta(\textbf{k}+\bar{\textbf{k}})}{(\bar{\omega}+\omega_{\textbf{k}}+i\epsilon)}-\frac{\delta(\textbf{k}+\bar{\textbf{k}})}{(\bar{\omega}-\omega_{\textbf{k}}-i\epsilon)}\right)\, e^{\frac{\pi\omega_{\textbf{k}}}{a}} +
\left(\frac{\delta(\textbf{k}+\bar{\textbf{k}})}{(\bar{\omega}-\omega_{\textbf{k}}+i\epsilon)}-\frac{\delta(\textbf{k}+\bar{\textbf{k}})}{(\bar{\omega}+\omega_{\textbf{k}}-i\epsilon)}\right)
	e^{-\frac{\pi\omega_{\textbf{k}}}{a}}\right]\right\}\\&=
	\frac{i\,n_{\beta_{0}}(\omega_{\bar{\textbf{k}}})}{2\sinh{\frac{\pi\omega_{\bar{\textbf{k}}}}{a}}}
	\left\{e^{\beta_{0}\omega_{\bar{\textbf{k}}}}\left[\frac{e^{\frac{\pi\omega_{\bar{\textbf{k}}}}{a}}}{(\bar{\omega}^{2}-\omega_{\bar{\textbf{k}}}^{2}+i\epsilon)}  -
	\frac{e^{-\frac{\pi\omega_{\bar{\textbf{k}}}}{a}}}{(\bar{\omega}^{2}-\omega_{\bar{\textbf{k}}}^{2}-i\epsilon)}\right]-\left[\frac{e^{\frac{\pi\omega_{\bar{\textbf{k}}}}{a}}}{(\bar{\omega}^{2}-\omega_{\bar{\textbf{k}}}^{2}-i\epsilon)}  -
	\frac{e^{-\frac{\pi\omega_{\bar{\textbf{k}}}}{a}}}{(\bar{\omega}^{2}-\omega_{\bar{\textbf{k}}}^{2}+i\epsilon)}\right]\right\}\,,
\end{aligned}\end{equation}
where the last line is obtained after performing the $\textbf{k}$-integral. Finally re-arranging the above we obtain (\ref{TF38}).

{\underline{$G_{LL}$}}: --
Use the expression Eq. (\ref{TF37}) and (\ref{TF38a}) along with (\ref{TFa2}) yields
\begin{equation}\label{TF}\begin{aligned}
iG_{LL} (\bar{k})= &\int_{-\infty}^{\infty}d\Delta\eta\int_{-\infty}^{\infty}d\Delta\xi\,iG_{{LL}}\left(\Delta\xi,\Delta\eta\right)\,e^{-i\bar{\textbf{k}}\Delta\xi+i\bar{\omega}\Delta\eta}\\&=
 i\int_{-\infty}^{\infty} \frac{d\textbf{k}~n_{\beta_{0}}(\omega_{\textbf{k}})}{4\omega_{\textbf{k}}\sinh{\frac{\pi\omega_{\textbf{k}}}{a}}}
 \\&      \left\{e^{\beta_{0}\omega_{\textbf{k}}}
 	\left[\left( \frac{\delta(\textbf{k}-\bar{\textbf{k}})}{(\bar{\omega}+\omega_{\textbf{k}}+i\epsilon)}
	-\frac{\delta(\textbf{k}-\bar{\textbf{k}})}{(\bar{\omega}-\omega_{\textbf{k}}-i\epsilon)}\right)\, e^{\frac{\pi\omega_{\textbf{k}}}{a}}
	 +
	\left(\frac{\delta(\textbf{k}-\bar{\textbf{k}})}{(\bar{\omega}-\omega_{\textbf{k}}+i\epsilon)}
	-\frac{\delta(\textbf{k}-\bar{\textbf{k}})}{(\bar{\omega}+\omega_{\textbf{k}}-i\epsilon)}\right)
	e^{-\frac{\pi\omega_{\textbf{k}}}{a}}\right]\right.
	\\&+\left.\left[
	\left( \frac{\delta(\textbf{k}+\bar{\textbf{k}})}{(\bar{\omega}-\omega_{\textbf{k}}+i\epsilon)}-\frac{\delta(\textbf{k}+\bar{\textbf{k}})}{(\bar{\omega}+\omega_{\textbf{k}}-i\epsilon)}\right)\, e^{\frac{\pi\omega_{\textbf{k}}}{a}} 
	+
\left(\frac{\delta(\textbf{k}+\bar{\textbf{k}})}{(\bar{\omega}+\omega_{\textbf{k}}+i\epsilon)}-\frac{\delta(\textbf{k}+\bar{\textbf{k}})}{(\bar{\omega}-\omega_{\textbf{k}}-i\epsilon)}\right)
	e^{-\frac{\pi\omega_{\textbf{k}}}{a}}\right]\right\}
	\\&=
	\frac{i\,n_{\beta_{0}}(\omega_{\bar{\textbf{k}}})}{2\sinh{\frac{\pi\omega_{\bar{\textbf{k}}}}{a}}}
	\left\{e^{\beta_{0}\omega_{\bar{\textbf{k}}}}\left[
	-\frac{e^{\frac{\pi\omega_{\bar{\textbf{k}}}}{a}}}{(\bar{\omega}^{2}-\omega_{\bar{\textbf{k}}}^{2}-i\epsilon)}  +
	\frac{e^{-\frac{\pi\omega_{\bar{\textbf{k}}}}{a}}}{(\bar{\omega}^{2}-\omega_{\bar{\textbf{k}}}^{2}+i\epsilon)}\right]
	+
	\left[\frac{e^{\frac{\pi\omega_{\bar{\textbf{k}}}}{a}}}{(\bar{\omega}^{2}-\omega_{\bar{\textbf{k}}}^{2}+i\epsilon)}  -
	\frac{e^{-\frac{\pi\omega_{\bar{\textbf{k}}}}{a}}}{(\bar{\omega}^{2}-\omega_{\bar{\textbf{k}}}^{2}-i\epsilon)}\right]\right\}\,.
\end{aligned}\end{equation}
This reduces to (\ref{TF39}).

{\underline{$G_{LR}$}}: -- 
Similarly, we obtain
\begin{equation}\label{TF}\begin{aligned}
iG_{LR} (\bar{k})= &\int_{-\infty}^{\infty}d\Delta\eta\int_{-\infty}^{\infty}d\Delta\xi\,iG_{{LR}}\left(\Delta\xi,\Delta\eta\right)\,e^{-i\bar{\textbf{k}}\Delta\xi+i\bar{\omega}\Delta\eta}\\&=
 i\int_{-\infty}^{\infty} \frac{d\textbf{k}~n_{\beta_{0}}(\omega_{{\textbf{k}}})}{4\omega_{\textbf{k}}\sinh{\frac{\pi\omega_{\textbf{k}}}{a}}}
 \\&
	\left\{e^{\beta_{0}\omega_{{\textbf{k}}}}\left[\left( \frac{\delta(\textbf{k}-\bar{\textbf{k}})}{(\bar{\omega}-\omega_{\textbf{k}}+i\epsilon)}-\frac{\delta(\textbf{k}-\bar{\textbf{k}})}{(\bar{\omega}+\omega_{\textbf{k}}-i\epsilon)}\right)\,+
	\left(\frac{\delta(\textbf{k}-\bar{\textbf{k}})}{(\bar{\omega}+\omega_{\textbf{k}}+i\epsilon)}-\frac{\delta(\textbf{k}-\bar{\textbf{k}})}{(\bar{\omega}-\omega_{\textbf{k}}-i\epsilon)}\right)
\right]\right.\\&+
	\left.\left[\left( \frac{\delta(\textbf{k}+\bar{\textbf{k}})}{(\bar{\omega}+\omega_{\textbf{k}}+i\epsilon)}-\frac{\delta(\textbf{k}+\bar{\textbf{k}})}{(\bar{\omega}-\omega_{\textbf{k}}-i\epsilon)}\right)\,  +
\left(\frac{\delta(\textbf{k}+\bar{\textbf{k}})}{(\bar{\omega}-\omega_{\textbf{k}}+i\epsilon)}-\frac{\delta(\textbf{k}+\bar{\textbf{k}})}{(\bar{\omega}+\omega_{\textbf{k}}-i\epsilon)}\right)
\right]\right\}\\&=
	\frac{i\,n_{\beta_{0}}(\omega_{\bar{\textbf{k}}})}{2\sinh{\frac{\pi\omega_{\bar{\textbf{k}}}}{a}}}
	\left\{e^{\beta_{0}\omega_{\bar{\textbf{k}}}}\left[\frac{1}{(\bar{\omega}^{2}-\omega_{\bar{\textbf{k}}}^{2}+i\epsilon)}  -
	\frac{1}{(\bar{\omega}^{2}-\omega_{\bar{\textbf{k}}}^{2}-i\epsilon)}\right]-\left[\frac{1}{(\bar{\omega}^{2}-\omega_{\bar{\textbf{k}}}^{2}-i\epsilon)}  -
	\frac{1}{(\bar{\omega}^{2}-\omega_{\bar{\textbf{k}}}^{2}+i\epsilon)}\right]\right\} \,,
\end{aligned}\end{equation}
which finally leads to (\ref{TF40}).

Exactly in identical manner $G_{RL}(\bar{k})$ can be evaluated.

\end{appendix}
 
 \bibliography{bibtexfile}
\bibliographystyle{ieeetr}

\end{document}